\documentclass[aps,rmp,twocolumn,groupedaddress,floatfix]{revtex4-1}
\pdfoutput=1
\usepackage{natbib}
\usepackage{amsmath}
\usepackage{amsfonts}
\usepackage{amssymb}
\usepackage{graphicx} 

\def \M{M}
\def \F{F}
\def \V{{\sigma^2_k}}
\def \Mtilde{\tilde{\M}}
\def \Mvec{{\vec{\M}}}
\def \A{{\mathbf{A}}}
\def \pfix{{p_\mathrm{fix}}}
\def \heterozygosity{\pi}
\def \deltak{{\Delta k}}
\def \epsilonc{\epsilon_c}

\begin{document}

\title{Fluctuations in fitness distributions and the effects of weak linked selection on sequence evolution}
\author{Benjamin H. Good$^{1}$}
\author{Michael M. Desai$^{1}$}
\affiliation{\mbox{${}^1$Department of Organismic and Evolutionary Biology, Department of Physics, and} \mbox{FAS Center for Systems Biology, Harvard University}}

\begin{abstract}
Evolutionary dynamics and patterns of molecular evolution are strongly influenced by selection on linked regions of the genome, but our quantitative understanding of these effects remains incomplete.  Recent work has focused on predicting the distribution of fitness within an evolving population, and this forms the basis for several methods that leverage the fitness distribution to predict the patterns of genetic diversity when selection is strong. However, in weakly selected populations random fluctuations due to genetic drift are more severe, and neither the distribution of fitness nor the sequence diversity within the population are well understood.  Here, we briefly review the motivations behind the fitness-distribution picture, and summarize the general approaches that have been used to analyze this distribution in the strong-selection regime. We then extend these approaches to the case of weak selection, by outlining a perturbative treatment of selection at a large number of linked sites. This allows us to quantify the stochastic behavior of the fitness distribution and yields exact analytical predictions for the sequence diversity and substitution rate in the limit that selection is weak. 
\end{abstract}

\date{\today}

\maketitle

















A central goal of modern population genetics is to predict the diversity and fate of DNA sequences within a population, taking into account the joint effects of mutation, recombination, natural selection, and demography at the sequence level. Diversity is a fundamental feature on these genomic scales, since the typical mutation rates in most organisms are sufficiently large that a number of sequence variants are likely to coexist within the population at any given time \citep{lewontin:hubby:1966, kreitman:1983, begun:etal:2007, nik-zainal:etal:2012, nelson:etal:2012, rambaut:etal:2008}. It is therefore imperative that our models of sequence evolution should be able to describe a large number of variants at disparate sites within the genome, possibly with different effects on the reproductive fitness of each individual \citep{hahn:2008}. 

This picture of extensive diversity at the sequence level stands in contrast to the large body of population genetics theory developed during the first half of the 20th century, which typically focused on the fate of a single mutant allele (relative to the wildtype) at a single genetic locus. Numerous mathematical models have been proposed, even for this highly simplified scenario, which correspond to different underlying assumptions about the mechanisms and stochasticity of natural selection, the reproductive lifecycle of the organism, and so on \citep{ewens:2004}. In large populations, many of the differences between these various models become negligible, and an elegant theoretical description of the two-allele, single-locus system can be obtained from the standard diffusion limit \citep{kimura:1955}. The frequency $f$ of a mutant allele with fitness effect $s$ in a population of size $N$ is assumed to satisfy the stochastic differential equation
\begin{align}
\label{eq:diffusion}
\frac{\partial f}{\partial t} = \underbrace{s \left[ f(1-f) \right]}_{\text{selection}} + \underbrace{\sqrt{\frac{f(1-f)}{N}} \, \eta(t) }_{\text{genetic drift}}\, ,
\end{align}
where $\eta(t)$ is a stochastic noise term which will be defined in more detail below. Equation (\ref{eq:diffusion}) relates the rate of change in $f$ to the deterministic action of selection and the random effects of genetic drift, and it is formally equivalent to the diffusion equation for the probability distribution of $f$ typically cited in the population genetics literature \cite{korolev:etal:2010}. Although the full solution to Eq. (\ref{eq:diffusion}) is quite complicated \cite{kimura:1955, song:steinrucken:2012}, this diffusion model is simple enough to admit a number of useful and exact results, including the well-known formula for the probability of fixation of a new mutant 
\begin{align}
\label{eq:kimura-fixation}
\pfix = \frac{1-e^{-2s}}{1-e^{-2Ns}} \, , 
\end{align}
and the average pairwise heterozygosity
\begin{align}
\label{eq:prf-pi}
\heterozygosity = 2 \left( \frac{\mu}{s} \right) \left[ \frac{ e^{-2Ns}+2Ns-1}{1-e^{-2Ns}} \right] \, ,
\end{align}
in the limit of low mutation rate $\mu$. The historical impact of this diffusion model cannot be overstated, and these simple results played a large role in illuminating both the qualitative and quantitative effects of genetic drift arising from the finite size of the population. However, extending these single-locus results to an explicitly sequence-based setting proves to be quite challenging. 

In principle, one can treat the entire genome as a single locus with each possible genotype represented by a unique allele. A genome of length $L$ would therefore require $2^L$ separate alleles and a corresponding system of diffusion equations relating the $2^L-1$ independent allele frequencies. This clearly becomes unwieldy for large genomes since the number of alleles grows exponentially with $L$, and the sparse mutational connectivity between the different sequences and their varying fitnesses removes much of the desired symmetry from the problem \cite{ethier:kurtz:1987}. Even for a genome with just $L=2$ sites, exact solutions can only be found for a few special cases, and one must often resort to numerical calculations \cite{barton:etheridge:2004} or Monte-Carlo simulations \cite{hill:robertson:1966}.  

A popular alternative approach is to treat each site in the genome as a separate locus and assume some sort of quasi-independent evolution among the various loci, so that the single-locus model in Eq. (\ref{eq:diffusion}) applies to the \emph{marginal} nucleotide frequencies at each site \cite{sawyer:hartl:1992}. This independent-sites approximation, which is exact in the limit of infinite recombination, reflects a historical perception of  linkage as an infrequent and generally small correction to an otherwise freely-recombining set of loci, as is often the case for a quantitative trait with genetic contributions from several distant sites \cite{falconer:1960, barton:turelli:1991, neher:shraiman:qtl:2011}. But given the typical recombination rates in most organisms, this assumption is likely to break down on local genomic scales, and effectively asexual selection on particular haplotype blocks may be a more accurate description \cite{franklin:lewontin:1970, slatkin:1972}. Moreover, it has been shown that selection within these linked regions leads to large deviations from the predictions assuming independent evolution between the various sites, even after adjusting for possible reductions in the effective population size \cite{charlesworth:etal:1993, bustamante:etal:2001, comeron:kreitman:2002, messer:petrov:2012, good:desai:2012}. Correctly accounting for the effects of selection on local genomic scales remains one of the major outstanding problems in population genetics, and is a necessary prerequisite if we wish to take full advantage of the increasing availability of DNA sequence data in order to make inferences about the evolutionary forces acting within a population \cite{pool:etal:2010}.

Recent advances in this area have employed a third approach --- situated somewhere between the genotypes-as-alleles and sites-as-loci schemes --- in which the distribution of fitnesses in the population plays a central role \cite{ohta:kimura:1973, haigh:1978, tsimring:etal:1996, rouzine:etal:2003, desai:fisher:2007, park:krug:2007, neher:etal:2010, hallatschek:2011, goyal:etal:2012}. Although the fitness distribution may seem to be rather tangential to the sequence-oriented questions introduced above, this quantity turns out to play an important role in mediating the effects of linked selection within the population, and several promising methods  predict the behavior of individual sequences based on their interactions with this population-wide distribution \cite{hudson:kaplan:1994, ofallon:etal:2010, neher:shraiman:draft:2011, zeng:charlesworth:2011, walczak:etal:2012, good:etal:2012}. Instead of tracking the frequencies of all possible genotypes or just the marginal frequencies at each site, this approach requires an explicit model for the frequency of individuals at each possible fitness, otherwise known as a \emph{fitness-class}. Here too, the interactions between mutation, recombination, drift, and selection can be quite complex, and significant progress has been made only in the case where genetic drift is negligible compared to these other evolutionary forces. This can often be a reasonable approximation in many populations, since the effects of genetic drift are typically less severe for the fitness classes than for the frequencies of the underlying genotypes. 

Nevertheless, even in this fitness-class picture the effects of genetic drift cannot be excised completely, since they play a crucial role in the high-fitness ``nose'' of the fitness distribution that often controls the behavior in the rest of the population \cite{desai:fisher:2007, brunet:etal:2008, hallatschek:2011, goyal:etal:2012, neher:shraiman:2012}. Various ad-hoc methods have been devised to account for this drift-dominated nose and its relation to the deterministic behavior in the bulk population, which yield accurate predictions for simple quantities such as the average rate of adaptation and the fixation probability of new mutations. Yet because of their ad-hoc nature, it is not entirely clear when these approximations are likely to be valid, or whether they remain appropriate for more complicated quantities of interest. Furthermore, in populations with weaker selection this separation between the drift-dominated nose and the deterministic bulk starts to break down, and the random nature of genetic drift becomes important throughout the entire fitness distribution. 

In the present work, we follow an approach that is orthogonal to both the weak-drift limit of this fitness-class description as well as the weak-mutation limit implicit in the standard single-locus treatment. Rather, we seek a fitness-class description for a regime with weak selection at a large number of linked sites. Suitably defined, the neutral limit of the population ``fitness distribution" is exactly solvable, and the corrections in the presence of selection can be calculated order by order as a perturbation series in powers of the selection strength. The resulting expressions may have relevance to sequence data obtained from natural populations [particularly in the context of the nearly-neutral theory of evolution \cite{ohta:1992}], but their primary value is qualitative. The zeroth-order neutral description offers a valuable window into the stochastic aspects of the population fitness distribution in the absence of the complicating effects of selection, while the higher-order terms give the exact corrections from interference at a large number of linked sites and help illuminate the previously obscure transition to neutrality. The exact nature of these selective corrections provides a valuable check on a number of common heuristic assumptions in the literature, which should agree with our asymptotic results when selection becomes weak. 

\section{Fitness classes and the population fitness distribution}

The distribution of fitnesses within the population is itself a random object which changes in time and reflects the inherent stochasticity of the evolutionary process. Two populations with the same genetic composition and the same set of available mutations will typically possess different fitness distributions after evolving independently for the same amount of time, although these distributions will be related in some statistical sense. Like the stochastic frequency of a single mutant allele discussed above, the statistical properties of the fitness distribution can be described by a generalization of the diffusion model in Eq. (\ref{eq:diffusion}) that makes both the large population and long genome limits explicit. We consider a population of $N$ haploid individuals that acquire new mutations at a total rate $U$ per generation. We assume that these mutations occur over a large number of loci, each with relatively small contributions to the total fitness, so that a mutation of effect $s$ arising in an individual with (log) fitness $X$ increases its fitness to $X + s$. Furthermore, we assume that the number of loci is sufficiently large, and epistasis sufficiently weak, that the set of available mutations can be approximated by a continuous distribution of fitness effects $\rho(s)$ that remains constant throughout the relevant time interval. 

\begin{figure}
\centering
\includegraphics[width=0.95\columnwidth]{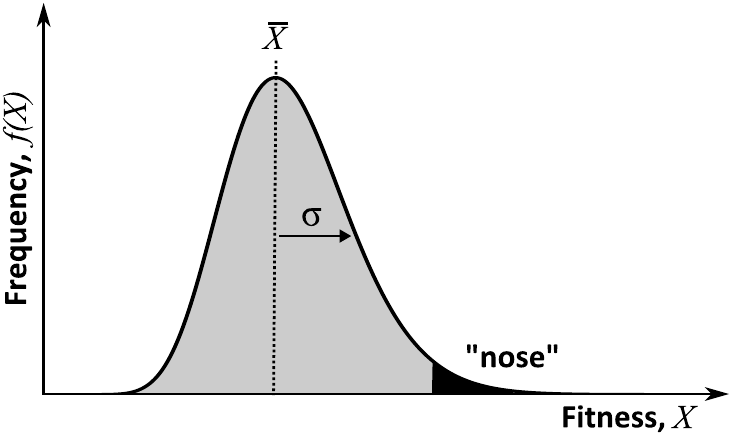}
\caption{A schematic depiction of the population fitness distribution, $f(X,t)$, which is obtained by grouping together genotypes with the same absolute fitness $X$. Important features of this distribution include the mean fitness $\overline{X}$ and the standard deviation $\sigma$, which is proportional to the typical fitness difference between two individuals in the population. We have also highlighted the high-fitness ``nose" of the distribution, where genetic drift continues to dominate even in extremely large populations. \label{fig:fitness-distribution}}
\end{figure}

The random arrival of new mutations and the effects of genetic drift are treated by a continuous-time stochastic model similar to the one introduced in \citet{hallatschek:2011}. Let $f(X,t)$ denote the relative frequency of individuals with absolute fitness $X$ at time $t$, normalized so that $\int dX \, f(X,t) = 1$. In some infinitesimal time $\delta t$, these frequencies are incremented according to the stochastic update rule 
\begin{equation}
\begin{aligned}
f(X,t+\delta t) & \propto f(X,t) + X f(X,t) \delta t \\ 
	& \quad + U \int ds \, \left[ f(X-s,t) - f(X,t) \right] \delta t  \\
	& \quad  + \sqrt{\frac{f(X,t) \delta t}{N}} \eta(X,t) \, ,
\end{aligned}
 \label{eq:pre-constraint-dynamics}
\end{equation}
where $\eta(X,t)$ denotes a set of independent Gaussian noise terms with zero mean and unit variance \cite{gardiner:1985}, and the constant of proportionality is chosen to satisfy the population size constraint
\begin{align}
\label{eq:population-size-constraint}
\int dX \, f(X,t+\delta t) = 1 \, .
\end{align}
This yields a familiar Langevin equation for the fitness distribution 
\begin{widetext}
\begin{equation}
\begin{aligned}
\frac{\partial f(X)}{\partial t} & = \underbrace{ \left[ X - \overline{X}(t) \right] f(X)}_\text{selection} + \underbrace{U \int ds \, \rho(s) \left[ f(X-s) - f(X) \right]}_\text{mutation}  +\underbrace{  \int dX' \, \left[ \delta(X'-X) - f(X) \right] \sqrt{\frac{f(X')}{N}} \eta(X')}_\text{genetic drift} \, , 
\end{aligned}
\label{eq:langevin}
\end{equation}
\end{widetext}
where $\overline{X}(t) = \int dX \, X f(X,t)$ is the mean fitness of the population (see Fig. \ref{fig:fitness-distribution}). Like the diffusion approximation at a single locus, this stochastic model is thought to describe the universal behavior that emerges in the limit that $N \to \infty$ and $L \to \infty$, while the per-site mutation rate $\mu$ and the relevant fitnesses $X$ tend to zero in such a way that the scaled quantities $N U \equiv NL\mu$ and $NX$ completely determine the dynamics. This scaling behavior provides an important check on our our intuition (as well as our algebra), since it implies that any effects that depend on $1/N$, $X$, or $N\mu$ alone are competely negligible in this model unless we explicitly relax one of these assumptions (e.g., the finite site effects in Appendix A).   

Two of the defining features of this stochastic model arise from the population size constraint in Eq. (\ref{eq:population-size-constraint}) that connects Eqs. (\ref{eq:pre-constraint-dynamics}) and (\ref{eq:langevin}): the resulting selection term becomes a nonlinear function of $f(X,t)$, and the previously simple noise terms acquire a complicated correlation structure. Such features are inherent in any model that imposes a population size constraint in this manner. The only fitnesses that matter are the relative fitnesses $x = X-\overline{X}(t)$, which depend not just on the properties of a particular DNA sequence, but also on the global behavior of all the sequences in the population. Moreover, the action of genetic drift is correlated among the various fitness classes in order to respect the constant population size.

Yet despite the complex correlation structure of this drift term, we have constructed our stochastic model so that its average effect vanishes at any particular time. Thus, we are led to examine the average profile $\langle f(X,t) \rangle$, which represents the expected value of the fitness distribution averaged over many independent populations. Taking the expectation of both sides of Eq. (\ref{eq:langevin}), we find that this average profile $\langle f(X,t) \rangle$ is governed by the \emph{deterministic} differential equation 
\begin{equation}
\begin{aligned}
\frac{\partial \langle f(X) \rangle}{\partial t} & = X \langle f(X) \rangle - \int dX' \, X' \langle f(X) f(X') \rangle \\
	& \quad \quad + \int ds \, \rho(s) \left[ \langle f(X-s) \rangle - \langle f(X) \rangle \right] \, .
\end{aligned}
\end{equation}
However, this moment equation does not close: the nonlinear selection term in Eq. (\ref{eq:langevin}) implies that the future behavior of $\langle f(X) \rangle$ does not just depend upon its own value in the present, but also on the two-point correlation function $\langle f(X) f(X') \rangle$. The time evolution of this two-point correlation function will in turn depend on the three-point correlation function, and so on. One must therefore solve an infinite hierarchy of these moment equations in order to obtain predictions for the mean behavior. A similar hierarchy arises for the single-locus diffusion in Eq. (\ref{eq:diffusion}), but in that case the simplicity of the drift term permits an exact solution. In contrast, the lack of closure among the moments of the fitness distribution is arguably the primary obstacle for a quantitative description of large numbers of interfering mutations, and many statistical properties of the fitness distribution remain unknown as a result. 

Much of the existing work in this field has essentially focused on various ways to approximate this correlated selection term. This is often achieved through some sort of approximate factorization of the form
\begin{align}
\label{eq:factorization}
\langle (X - \overline{X}) f(X) \rangle \approx (X-\langle \overline{X} \rangle) \langle f(X) \,  \rangle
\end{align}
so that the statistical aspects of the nonlinearity are marginalized\footnote{Strictly speaking, this sort of approximation is often more appropriate when this simple ensemble average is replaced with some other averaging scheme (see below) or an alternative measure of the ``typical'' behavior \cite{fisher:2012}}. Given that the magnitude of genetic drift is proportional to $1/N$, one regime where this approximation appears quite naturally is in the strong-selection limit $N|X-\overline{X}| \gg 1$, when the genetic drift term can be neglected in Eq. (\ref{eq:langevin}) and all higher correlations vanish. The canonical example of such a regime is the deleterious mutation-selection balance attained under strong purifying selection  \cite{haigh:1978}, which has been intensely studied in the context of Muller's ratchet \cite{muller:1964, stephan:etal:1993, gessler:1995, higgs:woodcock:1995, gordo:charlesworth:2000, etheridge:etal:2007, jain:2008, waxman:loewe:2010, neher:shraiman:2012} and background selection \cite{charlesworth:etal:1993, hudson:kaplan:1994, gordo:etal:2002, walczak:etal:2012, nicolaisen:desai:2012}. \citet{neher:shraiman:2012} have demonstrated that the deterministic limit becomes exact in this particular case when selection is infinitely strong. However, if beneficial mutations are present, or if some of the deleterious mutations are weakly selected, then even in this extreme limit the factorization in Eq. (\ref{eq:factorization}) does not hold for all $X$, since it starts to break down near the high-fitness ``nose'' of the distribution (see Fig. \ref{fig:fitness-distribution}), where a relatively small number of individuals have an outsized chance of taking over the population. Thus, in this strong-selection limit one often speaks of a division of the population into a drift-dominated nose (where stochasticity is extremely important) and a deterministic bulk where Eq. (\ref{eq:factorization}) holds. When the disinction between these regions is sufficiently sharp, a number of highly successful (although somewhat ad-hoc) approximations have been developed to treat the stochasticity in the nose and to self-consistently match this behavior with the deterministic bulk of the population \cite{tsimring:etal:1996, rouzine:etal:2003, desai:fisher:2007, goyal:etal:2012, neher:shraiman:2012}. Several alternative approaches are based on a modification of the stochastic dynamics in Eq. (\ref{eq:langevin}), which is chosen in a particular way so that the nonlinearity in the selection term vanishes by design \cite{hallatschek:2011, fisher:2012}. These models may be more appropriate when the boundary between the stochastic nose and the deterministic bulk is less pronounced, but their relation (and relevance) to the standard evolutionary model in Eq. (\ref{eq:langevin}) must be justified on an ad-hoc basis.

These methods are by far the most promising candidates for describing the evolutionary dynamics in a strong-selection regime relevant to many laboratory evolution experiments, microbial populations, or other rapidly adapting organisms. Yet from a purely theoretical standpoint, they suffer from a major shortcoming in that they attempt to describe a parameter regime for which no exact asymptotic description has been found. Although these methods were devised to approximate this asymptotic behavior, their correctness (apart from self-consistency) can only be validated by numerical comparisons to Monte-Carlo simulations of Eq. (\ref{eq:langevin}) for particular parameter values. This can make it difficult to test the individual assumptions that enter into these approximations or to compare different approximation methods, and it offers little direct information about which quantities or parameter regimes fall outside their domain of validity. On a more practical level, there may be many populations that are dominated by a large number of \emph{weakly} selected mutations where these strong-selection methods do not apply. In this case, few quantitative descriptions exist apart from assuming strict neutrality, and our knowledge of the relevant processes in this regime is extremeley limited. 

Thus, while previous approaches have attempted to reconcile the joint effects of selection and drift by mostly neglecting the latter, our approach here will be exactly the opposite. Rather than focus on those regimes where the selection term can be factored like Eq. (\ref{eq:factorization}), we consider the weak-selection limit where this selection term can be neglected entirely, or at least treated as a small perturbation. Unlike the strong-selection limit, the zeroth order solution in this nearly-neutral regime can be treated exactly and the full statistical behavior can be elucidated, which leads to a natural (and similarly exact) perturbation expansion in the presence of selection. 

\section{The neutral limit}

There are a variety of ways we could define the neutral limit of Eq. (\ref{eq:langevin}), but we are interested in one which does not lead to a trivial description of the resulting ``fitness distribution." For example, there is a naive limit in which the fitness effects of all new mutations have $s=0$, which implies that the entire population is confined to a single ``fitness class" with fitness $X=0$ for all time, i.e.
\begin{equation}
f(X) = \delta(X) \, .
\end{equation}
However, we can maintain much more of the interesting multi-locus behavior by ignoring the absolute fitness of each individual for the moment and concentrating instead on the number of mutations $k$ that each individual possesses. In this limit, a population-wide ``mutation number distribution'' $f(k,t)$ emerges in the same way that a fitness distribution $f(X,t)$ arises from Eq. (\ref{eq:langevin}). The stochastic dynamics in this case are governed by the Langevin equation
\begin{equation}
\begin{aligned}
\frac{\partial f(k)}{\partial t} & = U f(k-1) + U f(k) \\
	& \quad +  \sum_{k'} \left[ \delta_{k k'} - f(k) \right] \sqrt{\frac{f(k')}{N}} \eta(k',t) \, .
\end{aligned}
\label{eq:neutral-langevin}
\end{equation}
These dynamics are similar to the charge-ladder model introduced by \citet{ohta:kimura:1973}, which was initially created to model the early electrophoresis measurements of allelic diversity. This model later played an important role in the development of the neutral coalescent, which has since largely superseded it \cite{moran:1975, kingman:1976, kingman:1982,kingman:2000}. Our approach below will have much in common with this standard neutral result, although some quantities are more convenient to calculate in one framework than the other. However, our description in terms of fitness classes will lead to a natural generalization in the presence of selection, which is difficult to incorporate into the standard coalescent model \cite{neuhauser:krone:1997}. 

The stochastic dynamics in Eq. (\ref{eq:neutral-langevin}) are free of the nonlinearities that plagued our earlier analysis of Eq. (\ref{eq:langevin}), and the resulting equation for the average profile $\langle f(k,t) \rangle$ closes:
\begin{align}
\label{eq:neutral-deterministic}
\frac{\partial \langle f(k) \rangle}{\partial t} & = U \langle f(k-1) \rangle - U \langle f(k) \rangle \, .
\end{align}
This differential equation is straightforward to solve, and under the assumption that all individuals start with zero mutations at time $t=0$, we find that
\begin{equation}
\label{eq:average-solution}
\langle f(k,t) \rangle = \frac{(Ut)^k}{k!} e^{-Ut} \, .
\end{equation}
Thus, the mean of this distribution accumulates mutations at a constant rate $U$, which agrees with the standard calculation that assumes that each neutral mutation fixes independently. As $t \to \infty$, the width of this distribution grows larger and larger, and in order to conserve probability, $\langle f(k,t) \rangle$ approaches the trivial solution
\begin{align}
\label{eq:trivial-solution}
\lim_{t \to \infty} \langle f(k,t) \rangle = 0 \, .
\end{align} 
A similar observation was made previously in the context of the charge-ladder model, which reflects the fact that this earlier model and the one defined by Eq. (\ref{eq:neutral-langevin}) have no true stationary distribution. Intuitively, this degenerate behavior is an artifact of the averaging process we used in order to calculate $\langle f(k,t) \rangle$. While the average rate of mutation accumulation is simply the mutation rate $U$, the actual rate for any paticular population will tend to fluctuate around this value, and the location $\overline{k}(t)$ will become increasingly uncertain with time. By calculating the average $\langle f(k,t) \rangle$ as $t \to \infty$, we are effectively averaging many independent distributions whose centers are distributed across a large region of $k$, and hence the average number of individuals at any particular $k$ tends to zero. This line of reasoning is not specific to the neutral limit considered in this section, but is in fact a general property of any fitness distribution whose absolute location is subject to stochastic fluctuations. 

In all of these cases, the average distribution at long times is a poor summary of the typical distribution found in a random population. For example, while the width of the \emph{average distribution} in Eq. (\ref{eq:average-solution}) increases without bound, a simple argument from neutral coalescent theory shows the \emph{average width} of the population fitness distribution has a finite extent as $t \to \infty$. A random pair of individuals in the population will typically share a common ancestor $T_2 \sim N$ generations ago, so the difference between the number of mutations accumulated since the common ancestor is on the order of $NU$. We can see this in our current framework by simply measuring the number of mutations in each individual \emph{relative to the mean number of mutations in the population at any given time}. In particular, we can examine the variance in the number of mutations within the population, which is defined by
\begin{align}
\V = \sum_k (k - \overline{k})^2 f(k) \, .
\end{align}
Due to the presence of the $\overline{k}$ terms within this defintion, $\V$ is not just a simple linear function of the class sizes $f(k)$, and the rate of change of the average variance $\langle V \rangle$ cannot be written as a function of the average class sizes $\langle f(k) \rangle$ alone. Nevertheless, we can use the stochastic dynamics in Eq. (\ref{eq:neutral-langevin}) to show that the differential equation for $\langle \V \rangle$ does close on \emph{itself}, and we find that
\begin{align}
\frac{\partial \langle \V \rangle}{\partial t} & = U - \frac{\langle \V \rangle}{N} \, .
\end{align}
Again, assuming that all individuals start out with zero mutations at time $t=0$, this equation yields the simple solution
\begin{align}
\langle \V(t) \rangle = NU \left( 1 - e^{-t/N} \right) \approx \begin{cases} Ut & \text{if $t \ll N$,} \\
NU & \text{if $t \gg N$.} 
\end{cases}
\end{align}
Thus, we see that the variance attains an equilibrium value $\langle \V \rangle = NU$, as expected from our coalescent arguments, and it does so on the coalescent timescale $T_2 \sim N$. For $t \gg T_2$, the population continues to accumulate mutations at the same steady-state rate $U$, but it does so with the relatively constant shape dictated by this mutation-drift balance. On the other hand, for $t \ll T_2$ the average variance is essentially given by the deterministic estimate $Ut$ obtained from Eq. (\ref{eq:average-solution}). We argued earlier that this average distribution becomes unreliable when the uncertainty in the location of the mean becomes comparable to the width of a typical distribution. Our Langevin framework allows us to make this argument more explicit, since we can directly show that the variance in $\overline{k}$ obeys the differential equation 
\begin{equation}
\frac{\partial \mathrm{Var}(\overline{k})}{\partial t} = \frac{1}{N} \langle \V \rangle \, , 
\end{equation}
and hence
\begin{align}
\mathrm{Var}(\overline{k}) & = Ut - NU \left( 1 - e^{-t/N} \right) \\
	& \approx \begin{cases} \frac{Ut}{2} \left( \frac{t}{N} \right) & \text{if $t \ll N$} \\
N U \left( \frac{t}{N} \right) & \text{if $t \gg N$}
\end{cases}
\end{align}
Thus, when $t \sim N$, the uncertainty in $\overline{k}$ is on the order of the variance $\mu_2$ within a typical population, and the width of the average distribution $\langle f(k,t) \rangle$ will start to be dominated by the uncertainty in the mean. On much longer timescales, the distribution of mutations within a typical population will have a relatively tight width $\langle \V \rangle = NU$ and a mean $\overline{k}$ which moves deterministically towards higher mutation number at rate $U$, but which diffuses around this average position with diffusion constant $U$. 

\begin{figure}
\centering
\includegraphics[width=0.95\columnwidth]{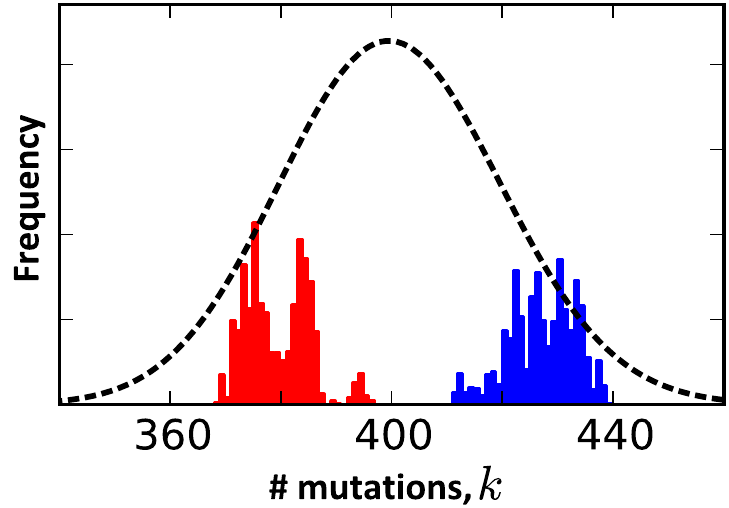}
\caption{The distribution of the number of neutral mutations within the population after $t=8N$ generations when $NU = 50$. The colored bars denote the results of two independent realizations of the stochastic dynamics in Eq. (\ref{eq:neutral-langevin}), and the dashed line is proportional to the average profile $\langle f_k(t) \rangle$ from Eq. (\ref{eq:average-solution}). [The vertical scale has been adjusted to improve visibility, and in reality all three distributions have the same total area.] \label{fig:neutral-fk}}
\end{figure}

\subsection{Higher moments and correlations}

Of course, the substitution rate and the steady-state variance $\langle \V  \rangle$ can be calculated by other means, without the need for the complicated machinery of the Langevin equation in Eq. (\ref{eq:neutral-langevin}). The real utility of this approach is that it allows us to calculate higher moments of the distribution of mutations that are inaccessible by these other methods. Motivated by our discussion of the variance in mutation number, we consider the family of \emph{central moments}, which are defined by 
\begin{align}
\M_m = \left\langle \sum_k (k-\overline{k})^m f(k) \right\rangle
\end{align}
for $m \geq 0$. We note that by definition, $\M_0 = 1$ and $\M_1 = 0$, while $\M_2$ is simply the variance $\langle \V \rangle$ discussed above. In Appendix C, we show that these central moments are governed by the compact equation
\begin{align}
\label{eq:general-moments-1}
\frac{\partial \M_m}{\partial \tau} & = NU \sum_{\ell=0}^{m-2} { m \choose \ell} \M_\ell - m \M_m \nonumber \\
	& \quad + {m \choose 2} \left\langle \sum_k (k-\overline{k})^2 f(k) \cdot \sum_k (k-\overline{k})^{m-2} f(k) \right\rangle \, ,
\end{align}
where we have rescaled time by $\tau = t/N$. Unfortunately, the nonlinear term on the right implies that these equations do not close when $m \geq 4$. We can only obtain a closed system by considering more complicated products of the form
\begin{align}
\label{eq:generalized-products}
M_{m_1,\ldots,m_J} = \left\langle \prod_{j=1}^{J} \left[ \sum_k (k-\overline{k})^{m_j} f(k,t) \right]  \right\rangle 
\end{align}
which have the general property that
\begin{align}
\M_{m,n} \neq \M_m \cdot \M_n \, .
\end{align}
The equations of motion for the first few moments $m \leq 4$ are relatively simple, and were first analyzed by \citet{higgs:woodcock:1995}. In our present notation, they showed that
\begin{align}
\frac{\partial \M_2}{\partial \tau} & = NU - \M_2 \label{eq:moment-eq-2} \\
\frac{\partial \M_3}{\partial \tau} & = NU - 3 \M_3 \\
\frac{\partial \M_4}{\partial \tau} & = NU + 6 NU \M_2 + 6 \M_{2,2} - 4 \M_4 \\
\frac{\partial \M_{2,2}}{\partial \tau} & = 2 N U \M_2 - 3 \M_{2,2} + \M_4
\end{align}
although one can technically include the fifth order moments
\begin{align}
\frac{\partial \M_5}{\partial \tau} & = NU + 10 NU \left( \M_2 + \M_3 \right) + 10 \M_{2,3} - 5 \M_5 \\
\frac{\partial \M_{2,3} }{\partial \tau} & = N U \left( \M_2 + \M_3 \right) - 8 \M_{2,3} + \M_5 \label{eq:moment-eq-23}
\end{align}
before triple products of the form $M_{m_1,m_2,m_3}$ start to appear. This system of first-order linear differential equations can be solved using standard Laplace transform methods, but in this case we are primarily interested in the steady-state behavior as $t \to \infty$. In this limit the time derivatives on the left-hand side vanish, and the resulting algebraic system can be easily solved to obtain 
\begin{align}
\M_2 & = NU \label{eq:moment-2} \\
\M_3 & = \frac{NU}{3} \\
\M_4 - 3 \M_{2,2} & = - 2 (NU)^2  \\
\M_{2,2} & = \frac{14 (NU)^2 + NU}{6} \label{eq:moment-22}
\end{align}
The first three of these quantities coincide with the first few cumulants of $f(k,t)$, which (through the related skew and kurtosis) are often used to characterize the shape of a distribution. However, since these are random distributions, we must be careful about the averaging process that we use to compute these characteristic quantities. For example, the excess kurtosis --- which is often used to measure the ``peakedness'' of the distribution and its deviations from normality --- could conceivably be calculated using any one of the four averages
\begin{align}
\frac{\M_4 - 3 \M_2^2}{\M_2^2} \, , \quad \frac{\M_4-3 \M_{2,2}}{\M_2^2} \, , \quad \frac{\M_4-3\M_{2,2}}{\M_{2,2}} \, , 
\end{align}
or
\begin{align}
\left\langle \frac{\left[ \sum_k (k-\overline{k})^4 f(k) \right] - 3 \left[ \sum_k (k-\overline{k})^2 f(k) \right]^2}{\left[ \sum_k (k-\overline{k})^2 f(k) \right]^2} \right\rangle \, ,
\end{align}
which each give slightly different results, even in the limit that $NU \to \infty$. One could argue that this last definition is closest to the standard usage of the excess kurtosis, but it is unfortunately the most difficult to calculate. Calculating the average of the inverse of a random variable typically requires us to first calculate all of its higher-order moments, which requires additional equations beyond Eqs. (\ref{eq:moment-2}-\ref{eq:moment-22}).

The second order product $M_{2,2}$ in Eq. (\ref{eq:moment-22}) can be used to calculate the variance in the width $\V$ between independent populations through the relation
\begin{align}
\mathrm{Var}(\V) = \M_{2,2} - \M_2^2 = \frac{8 (NU)^2 + NU}{6} \, .
\end{align}
This shows that the standard deviation in the typical variance at long times and large $NU$ is approximately
\begin{align}
\mathrm{Std}(\V) \sim 1.15 (NU) \, ,
\end{align}
which remains larger than its expected value $\langle \V \rangle = NU$ even as $NU\to\infty$. Thus, the variance in the number of mutations within the population is not self-averaging in the sense that $\sigma_k^2$ does not ``settle-down" to some fixed value in large populations. (One might naively expect this from the central limit theorem if the number of mutations in each individual was independent.) Instead, the typical spread in the number of mutations undergoes large fluctuations as the population continues to acquire new mutations. The typical lifetime of these fluctuations can be measured from the autocorrelation function
\begin{align}
G(\Delta \tau) & = \lim_{\tau \to \infty} \left[ \langle \V(\tau) \V(\tau+\Delta \tau) \rangle - \langle \V(\tau) \rangle \langle \V(\tau+\Delta \tau) \rangle \right] \nonumber \\
	& = \mathrm{Var}(\V) e^{-\Delta \tau}
\end{align}
which implies that these correlations decay in a simple manner on the coalescent timescale $T_2 \sim N$.

Continuing the system in Eqs. (\ref{eq:moment-eq-2}-\ref{eq:moment-eq-23}) to central moments with $m > 5$ starts to become complicated, since the moment equations for $\M_m$ start to involve more and more of the generalized products in Eq. (\ref{eq:generalized-products}). Nevertheless, the algebraic structure of these moment equations (which is derived in Appendix C) is such that the resulting system can be solved exactly in a straightforward manner with the help of a computer. The most important property of these moment equations is that the generalized products with $\sum m_j = m$ depend only on those products with total order less than or equal to $m$. Thus, at any given order we have a finite system of linear equations to solve. We can calculate these moments in an iterative manner. Given values for the generalized products at order $\leq m$, we can calculate the moments at order $m+1$ by solving the matrix equation
\begin{equation}
\label{eq:iterative-scheme}
\A_{m+1} \cdot \Mvec_{m+1} = \vec{b}(NU, \Mvec_1,\ldots,\Mvec_m) \, , 
\end{equation}
where $\Mvec_{m}$ is the collection of generalized products with order $\sum m_j = m$, $\mathbf{A}_m$ is a matrix of constants (independent of $NU$ or any of the moments), and $\vec{b}$ is a vector-valued function of $NU$ and the lower-order moments. The entries of the matrix $\A_m$, whose size is given by the number of generalized products at order $m$, can be determined directly by inspection from the system of equations in Appendix C. The resulting matrix must only be inverted once for each $m$, and then the analytical solutions for the various moments can be obtained by simple matrix multiplication. An implementation of this iterative algorithm in Python is available from the authors upon request.

\begin{figure}
\centering
\includegraphics[width=0.95\columnwidth]{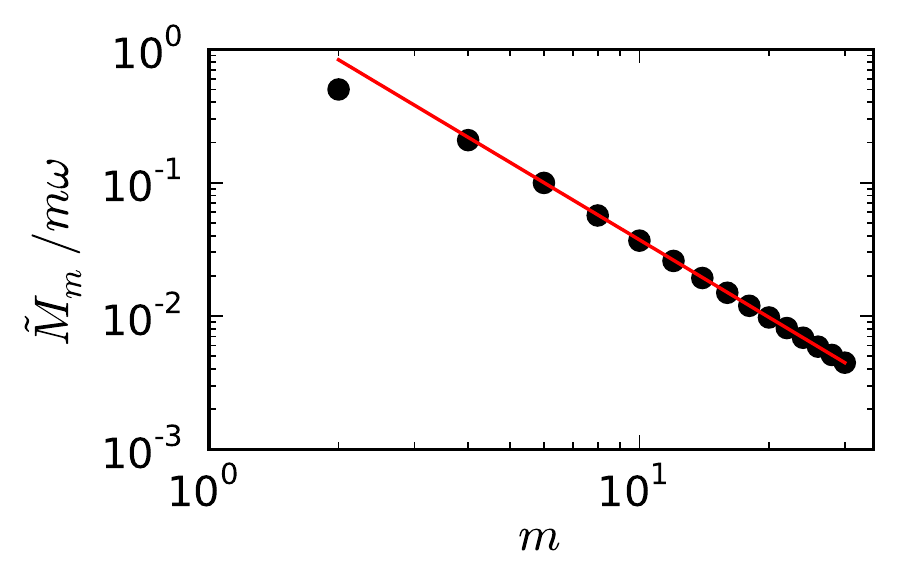}
\caption{The central moments $\M_m$ as a function of $m$ in the limit that $NU \to \infty$.  Symbols denote the exact numerical results calculated using the iterative scheme outlined in the text for $m=2,\ldots,30$. The red line denotes the approximate scaling form, $M_m \sim  m! (NU)^{m/2} / m^p$ where $p \approx 1.93$. \label{fig:neutral-moments}}
\end{figure}

One can in principle use this algorithm to calculate the moments for arbitrary $m$, which will be in the form of some polynomial in $NU$ similar to what we found for the first few moments in Eqs. (\ref{eq:moment-2}-\ref{eq:moment-22}). Typically, we will be interested in the limiting behavior for large $NU$, which we can access most easily by defining the rescaled moments
\begin{align}
\Mtilde_{m_1,\ldots,m_J} = \frac{\M_{m_1,\ldots,m_J}}{(\sqrt{NU})^{\sum m_j}} \, .
\end{align}
In the limit that $NU \to \infty$, the equations for the rescaled moments become independent of $NU$ and so $\Mtilde_m$ can depend only on $m$. These rescaled moments can be calculated from the same iterative scheme outlined above, and the results for the first thirty moments are shown in Fig. \ref{fig:neutral-moments}. For large $m$, these moments obey the approximate scaling relation 
\begin{align}
M_m \sim \left( \frac{m!}{m^p} \right) (NU)^{m/2}
\end{align}
where the exponent $p \approx 1.93$ can be extracted from the plot in Fig. \ref{fig:neutral-moments}. Although these central moments grow more quickly than the corresponding central moments of a Gaussian distribution, they grow sufficiently slowly that the centered distribution of mutations remains a light-tailed distribution. This is in contrast to the analogous neutral limit for Fisher-KPP waves, where the propagating front displays a power-law shape due to the periodic formation of smaller waves at the tip \cite{hallatschek:korolev:2009}. However, many of these technical details are beyond the scope of the present paper. The main point of this discussion for $m \geq 4$ is simply that \emph{all} of the central moments of this neutral distribution are \emph{exactly} solvable, as long as one is willing to devote the time and computing power necessary to implement the iterative scheme described above. 

\section{Perturbation theory for selected mutations}

Although some properties of the distribution of neutral mutations are interesting in their own right, previously developed methods like the neutral coalescent offer a simpler and more direct way to quantify the genetic diversity at the sequence level. Instead, the true utility of the exactly solvable model in the previous section lies in the fact that --- unlike these earlier methods --- our fitness-class description can be easily generalized to calculate the corrections that arise when selection is present. As an example, suppose the neutral mutations in the previous section now have a constant fitness effect $s$. Selection on these mutations leads to an additional term $+s(k-\overline{k})$ on the right-hand side of the stochastic dynamics in Eq. (\ref{eq:neutral-langevin}). Under these modified dynamics, the substitution rate $R = d\langle \overline{k} \rangle/dt$ for these mutations is now given by
\begin{align}
R = U + s \M_2
\end{align}
which now depends on the variance in $k$ in addition to the neutral accumulation rate $U$. The variance $\M_2$ will in turn depend upon the skew $\M_3$, and so on in the infinite hierarchy of moment equations mentioned earlier. However, if the strength of selection is weak and $Ns \ll 1$, the contribution to the variance from the $\M_3$ term will be small, and the variance will be approximately equal to the neutral result $\M_2 = N U$. Thus, by neglecting the selection term in the calculation of $\M_2$, we obtain an approximate expression for the substitution rate 
\begin{align}
R \approx U \left(1 + Ns \right)
\end{align}
valid in the limit that $Ns \to 0$ where the $Ns$ term is a small \emph{perturbative} correction to the neutral result $R = U$. In this way, the exact results for the neutral wave can serve as a basis for a perturbative analysis of the effects of selection in increasing powers of $Ns$. We note that this limit is quite distinct from the weak-selection regime analyzed by \citet{kimura:1955} for a single locus, and it is equally removed from the quasi-linked, weak-selection regime studied by \citet{nagylaki:1993}. By using the full multi-locus neutral limit as our starting point, we are able to analyze the selective corrections to all the sites simultaneously, while fully preserving the effects of linkage and interference found in the neutral case. Our analysis, which is anticipated to some extent in \citet{higgs:woodcock:1995}, is more similar to the perturbative treatment of noisy Fisher waves in \citet{hallatschek:korolev:2009}. 

In order for our perturbative scheme to apply to the more general populations described by Eq. (\ref{eq:langevin}), we must make some small modifications to our treatment of the neutral dynamics in order to properly account for distributions of fitness effects. In our analysis above, it was most natural to divide the population into fitness classes according to the discrete number of mutations $k$ in each individual. Now it will be convenient to consider $k$ to be a continuous variable, which is related to the fitness 
\begin{equation}
X = s k
\end{equation}
through an overall constant of proportionality $s$ that parameterizes the strength of selection. The  distribution of fitness effects $\rho(s)$ can then be alternatively viewed as a distribution of ``$k$-effects,'' which we denote by $\rho(\deltak)$. With these definitions, our model in Eq. (\ref{eq:langevin}) can be explicitly rewritten in ``$k$-space'' as
\begin{widetext}
\begin{equation}
\frac{\partial f(k)}{\partial \tau} = Ns ( k -\overline{k} ) f(k) + NU \int d(\deltak) \, \rho(\deltak) \left[ f(k-\deltak) - f(k) \right]  + \int dk' \, \left[ \delta(k'-k) - f(k) \right] \sqrt{f(k')} \eta(k') \, . \label{eq:k-langevin}
\end{equation}
\end{widetext}
where it is now clear that the nonlinear selection term is a perturbative correction with a well-defined limit when $Ns = 0$. Proceeding along the lines of the previous section, this stochastic differential equation yields an analogous set of differential equations for the central moments and the generalized moment products (see Appendix C).  The first few orders [which were obtained by \citet{higgs:woodcock:1995} and \citet{etheridge:etal:2007} for a similar model] are given by 
\begin{align}
\frac{\partial \M_2}{\partial t} & = NU \langle \deltak^2 \rangle - \M_2 + Ns \M_3 \label{eq:selected-moment-2} \\
\frac{\partial \M_3}{\partial t} & = NU \langle \deltak^3 \rangle - 3 \M_3 + Ns \langle \mu_4 - 3 \mu_2 \rangle \\
\frac{\partial \M_4}{\partial t} & = NU \langle \deltak^4 \rangle + 6 NU \langle \deltak^2 \rangle \M_2 \nonumber  \\ 
	 & \quad + 6 \M_{2,2} - 4 \M_4 + O(Ns) \\ 
\frac{\partial \M_{2,2}}{\partial \tau} & = 2 NU \langle \deltak^2 \rangle \M_2 +  \M_4 - 3 \M_{2,2} + O(Ns) \label{eq:selected-moment-22} 
\end{align}
where $\langle \deltak^m \rangle = \int d(\deltak) \, (\deltak)^p \rho(\deltak)$ denotes $m$th moment of the distribution of fitness effects. Thus, in the presence of selection the moments at order $m$ now include terms that depend on the moments at order $m+1$. The resulting system of equations cannot be solved at any fixed order because it always depends on the moments at a still-higher order. 

While this lack of closure among the moment equations makes it difficult to obtain a closed-form solution for any particular moment, it naturally suggests a perturbative approach similar to the $R \approx U(1+Ns)$ approximation above. In particular, we assume that in the limit $Ns \to 0$, each of the central moments $\M_m$ admits an asymptotic expansion of the form 
\begin{align}
\label{eq:asymptotic-expansion} 
\M_m \sim \sum_{j=0}^\infty \M^{(j)}_m (Ns)^j \, \quad (Ns \to 0) \, , 
\end{align}
where $\M^{(j)}_m$ is a numerical coefficient that depends only on $NU$. A similar expansion is assumed for the generalized products in Eq. (\ref{eq:generalized-products}). Substituting these expressions into the moment equations and grouping terms in powers of $Ns$, we can obtain a generalized system of equations for the coefficients $\M_m^{(j)}$, which is listed in Appendix C. The important feature of these equations is that unlike the case for the full moments $\M_m$, the equations for the coefficients $\M_m^{(j)}$ close for a given order $j$ and $m$. The algebraic properties of these equations are again sufficiently simple that they only lead to a slightly more complicated version of Eq. (\ref{eq:iterative-scheme}),
\begin{align}
\A_{m+1} \cdot \Mvec_{m+1}^{(j)} = \vec{b}\left( NU,\Mvec_1^{(j)},\ldots,\Mvec_m^{(j)} \right) + \vec{c}\left( \Mvec^{(j-1)}_{m+1} \right)
\end{align}
where $\A$ and $\vec{b}$ are the same as in the neutral case, and $\vec{c}$ is a vector-valued function of the moments at the next-lowest order in $j$. Thus, the iterative procedure outlined for the neutral case can be easily generalized to calculate the coefficients $\Mvec^{(j)}_m$ order-by-order for arbitrary $m$ and $j$. An implementation written in Python is available from the authors upon request. 

\begin{figure}
\centering
\includegraphics[width=0.95\columnwidth]{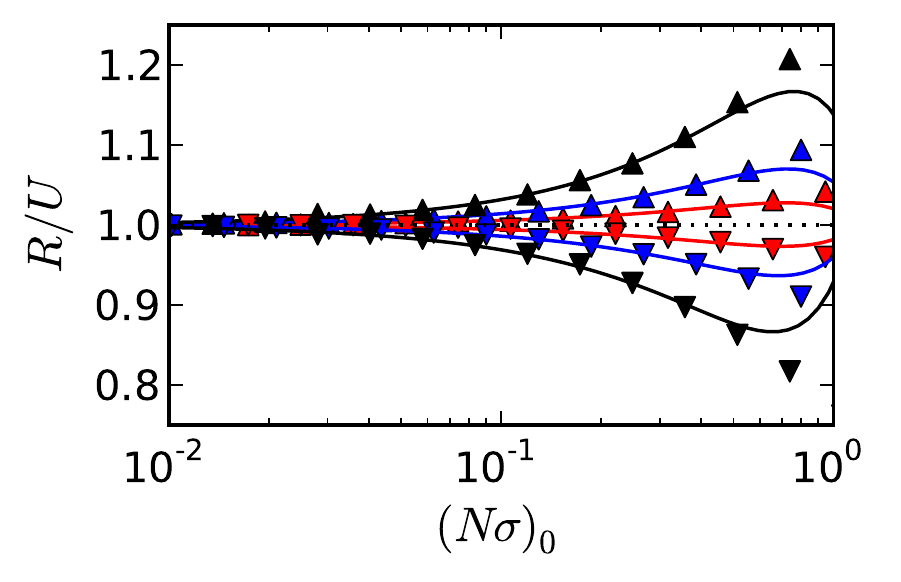}
\caption{The scaled substitution rate $R$ as a function of the zeroth-order fitness variance $(N \sigma)^2 = NU(Ns)^2$. Symbols denote the results of forward-time simulations for $NU=10$ (black), $NU=50$ (blue), and $NU=300$ (red), and the solid lines give the predictions from Eq. (\ref{eq:velocity}). Upper triangles denote populations with a purely beneficial distribution of fitness effects $\rho(\Delta k) = \delta(\Delta k - 1)$, while the lower triangles give the corresponding deleterious distribution $\rho(\Delta k) = \delta(\Delta k + 1)$. Our predictions start to diverge near $N\sigma=1$, when we expect our perturbation expansion to break down.  \label{fig:substitution-rate}}
\end{figure}

As an alternative to this explicit order-by-order calculation, we note that the coefficients in the asymptotic expansion in Eq. (\ref{eq:asymptotic-expansion}) are unique \cite{hinch:1991}, so we can also obtain these coefficients simply by dropping the selection term in the moment hierarchy at the desired order, solving the resulting finite system of equations, and then reexpanding the solution in powers of $Ns$. Applying this procedure to the moment equations in Eqs. (\ref{eq:selected-moment-2}-\ref{eq:selected-moment-22}), we obtain the first few corrections to the population variance in fitness
\begin{equation} 
\begin{aligned}
\M_2 & = N U \langle \deltak^2 \rangle + \left( \frac{NU \langle \deltak^3 \rangle}{3} \right) Ns \\
	& \quad \quad  - \left( \frac{2 [NU \langle \deltak^2 \rangle]^2}{3} \right) (Ns)^2 + O(Ns)^3 
\end{aligned}
\end{equation}
and hence the rate of adaptation $v = \frac{d \langle \overline{X} \rangle}{dt}$ is given by 
\begin{equation}
\begin{aligned}
v & = Us \left[ \langle \deltak \rangle + Ns \langle \deltak^2 \rangle + \frac{(Ns)^2 \langle \deltak^3 \rangle}{3} \right. \\
	& \quad \quad \quad \left. - \frac{2 [NU \langle \deltak^2 \rangle (Ns)^2][Ns \langle \deltak^2 \rangle]}{3} \right] + O(Ns)^4 
\end{aligned}
\label{eq:velocity}
\end{equation}
The first three terms in this expansion are exactly what one would obtain by assuming that the individual sites evolve independently, in which case the rate of fitness increase would simply be the sum of the single-locus adaptation rates that can be calculated from Eq. (\ref{eq:kimura-fixation}). The fourth term in this expansion represents a fundamentally new correction that arises solely from the accumulation of selected mutations over many different sites. This term is proportional to the variance in fitness within the population, so we see that the rate of adaptation is reduced in populations with a larger number of selected mutations (see Fig. \ref{fig:substitution-rate}) because many of these mutations will be lost to clonal interference before they can fix. Similarly, in populations which are accumulating deleterious mutations due to Muller's ratchet, this interference term leads to an increase in the rate of Muller's ratchet (again, see Fig. \ref{fig:substitution-rate}) due to the increased importance of fluctuations in the high-fitness nose of the population. 

In addition to the mean rate of adaptation, we can use this perturbative scheme to calculate the fluctuations in the rate of adaptation as well, which has so far been accessible only through heuristic arguments. Using the dynamics in Eq. (\ref{eq:k-langevin}), we can construct similar moment hierarchy for $\mathrm{Var}(\overline{k})$ and its relatives, and the first few orders are given by 
\begin{align}
\frac{\partial \mathrm{Var}(\overline{k})}{\partial \tau} & = \M_2 + 2 Ns \mathrm{Cov}(\overline{k},\M_2) \\
\frac{\partial \mathrm{Cov}(\overline{k},M_2)}{\partial \tau} & = - \mathrm{Cov}(\overline{k},\M_2) + \M_3 \nonumber \\
	& \quad + Ns \left[ \mathrm{Var}(\M_2) + \mathrm{Cov}(\overline{k},\M_3) \right] \\
\frac{\partial \mathrm{Cov}(\overline{k},\M_3)}{\partial \tau} & = - 3 \mathrm{Cov}(\overline{k},\M_3) + \M_4 -3 \M_{2,2} + O(Ns)
\end{align}
Thus, at long times $t \to \infty$, the mean fitness fluctuates diffusively
\begin{align}
\mathrm{Var}(\overline{k}) \sim 2 D t \quad (t \to \infty) \, , 
\end{align}
with diffusion constant
\begin{equation}
\begin{aligned}
\label{eq:diffusion-constant}
D & = \frac{U \langle \deltak^2\rangle}{2} \left[ 1 +  \frac{\langle \deltak^3\rangle}{\langle \deltak^2 \rangle}Ns - \frac{2 NU \langle \deltak^2 \rangle (Ns)^2}{3} \right. \\
	& \quad \quad \quad \quad \quad \quad \left. + \frac{NU \langle \deltak^4 \rangle (Ns)^2}{3}  \right] + O(Ns)^3 \, .
\end{aligned}
\end{equation}
Again, we see that interference between the various lineages leads to a reduction in the diffusivity of the mean fitness that is proportional to the variance in fitness within the population. This leads to a novel prediction in the case of the dynamic mutation-selection balance discussed in \citet{goyal:etal:2012}, where a balance between beneficial and deleterious substitutions halts any global fitness change in the population. The results in Eq. (\ref{eq:diffusion-constant}) show that even though the average fitness change in these populations is zero, we still expect the distribution to wander diffusively around this fixed point, with diffusion constant
\begin{align}
D(\epsilonc) = \frac{U}{2} \left[ 1 +  Ns(1-2\epsilonc) - \frac{NU (Ns)^2}{3} \right] + O(Ns)^3
\end{align} 
in the limit that $Ns \to 0$, where $\epsilonc$ is the critical ratio of beneficial to deleterious mutations. This scenario is discussed in more detail in Appendix A.

\section{Application to sequence evolution}

Our analysis so far has focused on population-wide properties of the fitness distribution and important aspects of the evolutionary dynamics such as the rate of adaptation. In the introduction however, we were primarily interested in predicting evolutionary fates and diversity at the sequence level, which we have so far neglected. In the present section, we will demonstrate how this fitness-class description can also provide a window into the evolutionary dynamics of sequences \emph{within} a particular population. 

\subsection{Fate of a focal lineage}

As an example, we consider the fate of some clonal lineage within the population which has an initial frequency $p$ and fitness $X_0 = s k_0$ at time $t=0$. This lineage could consist of a group of individuals that share a common point mutation at a particular site or it could alternatively represent some fluorescently labeled ``marker population'' whose dynamics we wish to follow. Now in addition to tracking the fitness of each individual in the population, we must also keep track of whether these individuals are descended from this particular lineage or from the background population. We can accomplish this by dividing the occupation densities $f(k,t)$ into two classes $f_1(k,t)$ and $f_0(k,t)$ which contain individuals descended from the focal lineage and the background population respectively. This division requires a small modification to our original dynamics in Eq. (\ref{eq:k-langevin}), 
\begin{widetext}
\begin{align}
\label{eq:marker-langevin}
\frac{\partial f_i(k)}{\partial \tau} & = Ns ( k - \overline{k} ) f_i(k) + NU \int d(\deltak) \rho(\deltak) \left[ f_i(k-\deltak) - f(k) \right] + \sum_{k',j} \left[ \delta(k'-k) \delta_{ij} - f_i(k) \right] \sqrt{f_j(k)} \eta_j(k) \, ,
\end{align}
\end{widetext}
but otherwise the stochastic dynamics are essentially the same. Now in addition to the population-wide central moments $\M_m$, we can also introduce a new set of central moments 
\begin{align}
\F_m = \left\langle \int dk \, (k-\overline{k})^m f_1(k,t) \right\rangle \, ,
\end{align}
that are specific to the focal lineage. The lowest-order moment $\F_0(t)$ represents the average total fraction of the population that is descended from the focal lineage (e.g., the frequency of a particular SNP within the population). At long times, one of two things can happen: either the focal lineage is outcompeted by the background and $\int dk \, f_1(k,t) = 0$, or its descendants take over the population and $\int dk \, f_1(k,t) = 1$. Thus, the probability of fixation is given by
\begin{equation}
p_\mathrm{fix} = \lim_{t \to \infty} F_0(t) \, .
\end{equation}
Using the dynamics in Eq. (\ref{eq:marker-langevin}) and the rules of our stochastic calculus, we obtain an additional moment hierarchy for the focal moments $\F_m$. As we show in Appendix D, this requires us to consider generalized moments of the form
\begin{equation}
\begin{aligned}
F_{\vec{m};\vec{n}} & = \left\langle \prod_{j=1}^{J_m} \left[ \int dk \, (k-\overline{k})^{m_j} f_1(k,t) \right] \right. \\
	& \quad \quad \times \left. \prod_{j=1}^{J_n} \left[ \sum_i \int dk \, (k-\overline{k})^{n_j} f_i(k,t) \right] \right\rangle
\end{aligned}
\label{eq:generalized-focal-products}
\end{equation}
where the $m$ indices denote moments specific to the focal lineage and the $n$ indices denote the population-wide moments $\M_{m}$ discussed earlier. In this notation, the first few orders of the moment hierarchy for $\F_0$ are given by 
\begin{align}
\frac{\partial \F_0}{\partial \tau} & = Ns \F_1 \\
\frac{\partial \F_1}{\partial \tau} & = - \F_1 + Ns  \left( \F_2 - \F_{0;2} \right) \\
\frac{\partial \F_2}{\partial \tau} & = N U \langle \deltak^2 \rangle \F_0 + \F_{0;2} - 2 \F_2 + Ns \left( \F_3 - 2 \F_{1;2} \right) \\
\frac{\partial \F_{0;2} }{\partial \tau} & = NU \langle \deltak^2 \rangle \F_0 + \F_2 - 2 \F_{0;2} + Ns \left( \F_{1;2} + \F_{0;3} \right) \\
\frac{\partial \F_3}{\partial \tau} & = N U \langle \deltak^3 \rangle \F_0 + 3 N U \langle \deltak^2 \rangle \F_1 \nonumber \\
	& \quad \quad - 3 \F_3 + 3 \F_{1;2} + O(Ns) \\ 
\frac{\partial \F_{0;3}}{\partial \tau} & = N U \langle \deltak^3\rangle \F_0 + \F_3 - 4 \F_{0;3}  - 3 \F_{1;2} + O(Ns) \\
\frac{\partial \F_{1;2}}{\partial \tau} & = NU \langle \deltak^2 \rangle \F_1 + \F_3 - \F_{0;3} - 3 \F_{1;2} + O(Ns)
\end{align} 
We are interested in the limiting behavior at long times, which is most easily obtained via the Laplace transformed moments 
\begin{equation}
\F_{\vec{m};\vec{n}}(z) = \int e^{-z \tau} \F_{\vec{m};\vec{n}}(\tau) \, d\tau \, ,
\end{equation}
which satisfy the identity
\begin{align}
\F_0(\tau) \sim \F_0(0) + Ns \cdot \F_1(z=1/\tau) \, ,
\end{align}
as $\tau \to \infty$. Taking the Laplace transform of the first few equations in the moment hierarchy and truncating the selection terms at order $O(Ns)^4$, we can immediately conclude that 
\begin{align}
p_{\mathrm{fix}} & = \F_0(0) + Ns \F_1(0) + \frac{(Ns)^2}{3} \left[ \F_2(0) - \F_{0;2}(0) \right] \nonumber \\
	& \quad \quad - \frac{(Ns)^3}{3} \left[ \F_{1;2}(0) + NU \langle \deltak^2 \rangle \right] + O(Ns)^4 \, .
\end{align}
The initial conditions for the various moments are given by 
\begin{align}
\F_m(0) & = p(1-p)^m \left[ k_0 - \left( \overline{k} \right)_\mathrm{bg} \right]^m \\
\M_m(0) & = \F_m(0) + (1-p) \sum_{\ell=0}^m {m \choose \ell} (-p)^\ell \left[ k_0 - (\overline{k})_\mathrm{bg} \right]^\ell \nonumber \\
	& \quad \times  (\M_{m-\ell})_\mathrm{bg}
\end{align}
where 
\begin{align}
(\overline{k})_\mathrm{bg} & = \sum_k k f_0(k,0) \, , \\
(\M_m)_\mathrm{bg} & = \sum_k [k-(\overline{k})_{\mathrm{bg}}]^m f_0(k,0)
\end{align}
denote the mean fitness and the central moments of the background popuation, respectively. For a lineage created by a spontaneous mutation, the initial frequency is just $p = 1/N$. The initial fitness $s k_0$ is obtained as a random draw from the population fitness distribution plus the fitness effect $s \deltak$ of the mutation. After averaging over the possible fitness backgrounds that this mutation could have arisen on, we find that  
\begin{align}
[k_0 - (\overline{k})_\mathrm{bg}]^m \to \sum_{\ell=0}^m {m \choose \ell} (\Delta k)^\ell (M_{m-\ell})_\mathrm{bg} 
\end{align}
Averaging over the background moments $(\M_m)_\mathrm{bg}$ simply yields the steady-state moments that we derived in the previous section. Thus, in the limit of large population sizes we have
\begin{align}
\F_m(0) & = \frac{1}{N} \sum_{\ell=0}^m {m \choose \ell} (\deltak)^\ell \M_{m-\ell} \\
\M_m(0) & = \M_m
\end{align}
and the fixation probability for a spontaneous mutation with effect $\deltak$ is given by
\begin{align}
\label{eq:fixation-probability}
p_\mathrm{fix}(\deltak) & = \frac{1}{N} \left[ 1 + Ns \deltak + \frac{(Ns \deltak)^2}{3}\right. \nonumber \\
	& \quad \quad\left. - \frac{2 [NU \langle \deltak^2 \rangle (Ns)^2] [Ns \deltak]}{3} \right] + O(Ns)^4 
\end{align} 
Again, we see that the first three terms are identical to the single-locus result in Eq. (\ref{eq:kimura-fixation}). We obtain the lowest-order ``interference correction" in the fourth term, which reduces the probability of fixation of a benficial mutation in a way that is directly proportional to the average variance in fitness within the population at the time of the mutation. For a deleterious mutation, this correction term actually increases the fixation probability because the mutant could find itself on an anomalously fit background (thus mitigating some of the effect of the deleterious mutation). We recover the standard neutral fixation probability $p_\mathrm{fix} = 1/N$ when $\deltak = 0$.

Noting the similarity between the fixation probability in Eq. (\ref{eq:fixation-probability}) and the rate of adaptation in Eq. (\ref{eq:velocity}), we see that the relation
\begin{align}
v = \int N U \cdot s \deltak \cdot \pfix(\deltak) \cdot \rho(\deltak) \, d(\deltak)
\end{align}  
holds at least through the first few orders in $Ns$. This relation has formed the basis for several studies of the evolutionary dynamics under strong selection \cite{neher:etal:2010, hallatschek:2011, good:etal:2012}, with the additional ``mean-field" ansatz
\begin{align}
\label{eq:mean-field-anzatz}
\pfix(\deltak) \approx \int dx \, \langle f(x-s\deltak) \rangle \cdot \pfix(x|\langle f(x) \rangle)  \, .
\end{align}
Here, $\pfix(x|\langle f(x) \rangle)$ denotes the fixation probability for a new mutation with relative fitness $x$, given that the centered fitness distribution of the rest of the population is $\langle f(x) \rangle$. However, we see that in the present regime, this mean-field ansatz is not quite correct. Instead, we require the slightly more complicated average 
\begin{align}
\label{eq:actual-average}
\pfix(\deltak) \approx \int dx \, \langle f(x-s\deltak) \cdot \pfix(x|f(x)) \rangle  \, ,
\end{align}
which jointly considers the fluctuations in the fitness background of the mutant as well as the fluctuations in the fitnesses of its competing lineages. Like the other correlated quantities we have considered in the present work, this average more or less decouples in the strong selection limit $Ns \to \infty$, and we recover the ``mean-field'' ansatz in Eq. (\ref{eq:mean-field-anzatz}). But in the weak selection regime considered here, these correlated fluctuations start to become more important, and Eq. (\ref{eq:actual-average}) is required in order to correctly account for the population-level dynamics.

\subsection{Diversity at a focal site}

In addition to predicting the ultimate fate of a sequence, these focal lineage dynamics can also be used to predict the average heterozygosity at a particular site along the genome and therefore the overall levels of sequence diversity in the population. We consider a particular site within the genome with a per-site mutation rate $\mu$ and scaled fitness effect $\deltak$. This site will be polymorphic in a randomly sampled pair of individuals if and only if (1) this site mutated at some time $t$ in the past and (2) exactly one member of the pair was drawn from the mutant lineage, which has size $\int dk \, f_1(k,t)$ in the present. After averaging overall possible mutation times (and taking note of the fact that the backward-time mutation process is Poisson), we find that 
\begin{equation}
\begin{aligned}
\heterozygosity & =  \left\langle \int_0^\infty dt \, N\mu e^{-N\mu t} \right. \\
	& \quad \quad \times \left. 2 \left( \int dk \, f_1(k) \right) \left( 1-\int dk \, f_1(k) \right)  \right\rangle
\end{aligned}
\end{equation}
In the infinite-sites limit where $N \mu \to 0$, this yields the relation
\begin{align}
\heterozygosity = 2 N \mu \int_0^\infty N H(\tau) d\tau
\end{align}
where we have defined the heterozygosity function
\begin{align}
H(\tau) & = \left\langle \left( \int dk \, f_1(k,t) \right) \left(1-\int dk \, f_1(k,t) \right) \right\rangle \nonumber \\
	& = F_0(\tau) - F_{0,0}(\tau)
\end{align}
Again, we can use the dynamics in Eq. (\ref{eq:marker-langevin}) to construct a similar hierarchy of moment equations for the heterozygosity, and the first few orders are given by  
\begin{align}
\frac{\partial H}{\partial \tau} & = - H + Ns \left( \F_1 - 2 \F_{0,1} \right) \\
\frac{\partial (\F_1 - 2 \F_{0,1} )}{\partial \tau} & = - 3 \left( \F_1 - 2 \F_{0,1} \right) + Ns \left( \F_2 - 2 \F_{0,2} - \F_{0;2} \right) \nonumber \\
	& \quad \quad  + Ns \left( 2 \F_{0,0;2} - 2 \F_{1,1} \right) \\
\frac{\partial \F_{0,2}}{\partial \tau} & = N U \langle \deltak^2 \rangle \left( \F_0 - H \right) + \F_2 + \F_{0,0;2} \nonumber \\
	& \quad \quad - 3 F_{0,2} - 2 \F_{1,1} + O(Ns) \\
\frac{\partial \F_{0,0;2}}{\partial \tau} & =  N U \langle \deltak^2 \rangle \left( \F_0 - H \right) + \F_{0;2} \nonumber \\
	& \quad \quad - 4 \F_{0,0;2} + 2 \F_{0,2} + O(Ns) \\
\frac{\partial \F_{1,1} }{\partial \tau} & = - 3 \F_{1,1} + \F_2 - 2 \F_{0,2} + \F_{0,0;2} + O(Ns)
\end{align} 
Truncating the hierarchy and solving the Laplace transformed equations, we obtain 
\begin{align}
\heterozygosity = 2 N \mu \left[ 1 + \frac{Ns \deltak}{3} - \frac{2 NU \langle \deltak^2 \rangle (Ns)^2}{9} \right] + O(Ns)^3 
\end{align}
Again, the first two terms in this expansion are equivalent to the single-locus results in Eq. (\ref{eq:prf-pi}). At third order in $Ns$, we obtain the lowest-order correction due to interference at neighboring sites, which reduces the diversity at a particular site according to the variance in fitness within the population. This reduction in diversity is seen even at putatively neutral or synonymous sites that are not otherwise selected on their own. 

\section{Discussion}

Although natural selection acts on the genome as a whole, the effects of selection at a large number of linked sites are only beginning to be characterized. Recent studies have identified the distribution of fitnesses within the population as a key mediator for these effects, but our understanding of this distribution remains limited to a few special cases where the strength of selection is strong and genetic drift is correspondingly weak. Here, we have introduced a general method for analyzing the effects of selection at many linked loci, which incorporates linkage and drift exactly while treating the global strength of selection as a perturbative correction. This framework allows us to investigate the stochastic behavior of the fitness distribution in a regime where the fluctuations due to drift are especially strong, and it fills an important gap in our theoretical understanding of linked selection in the approach to the neutral limit. 

As a quantitative theory, the present framework suffers from several shortcomings that may limit its direct applicability to data from natural populations. Our perturbative approach gives predictions for various quantities in terms of an asymptotic series in the limit that $Ns \to 0$, which means that for a fixed number of terms in this series, the resulting formulae will only become valid once $Ns$ is sufficiently small. Moreover, these aysmptotic expressions are \emph{nonuniform} as a function of the mutation rate $NU$, and in general for larger mutation rates we require ever smaller values of $Ns$ for our expressions to remain accurate. In reality, these asymptotic series could be more accurately described as an expansion in powers of the typical fitness variance $N\sigma \approx Ns \sqrt{NU}$, valid in the limit that $N\sigma \lesssim 1$. 

It remains an open question exactly what values of $N \sigma$ are relevant for natural populations. Indeed, one of the major motivating factors behind this quantitative approach to linked selection is to eventually use these theoretical tools to infer $N \sigma$ directly from DNA polymorphism data in sampled populations. Because the vast majority of new mutations are thought to be either neutral or weakly deleterious, there has been speculation that evolution at the sequence level is dominated by these ``nearly-neutral'' mutations with $Ns \lesssim 1$ \cite{ohta:1992}, although it is unclear whether these selection coefficients lead to an $N \sigma$ that is sufficiently small for our results to apply. In principle, the range of applicability of our expressions can be improved by including more terms in the expansion, but there is typically an upper limit to the radius of convergence that can be achieved this way \cite{hinch:1991}. However, because we have outlined a method for calculating successively higher-order terms programatically, series improvement methods could potentially be used to extend the radius of convergence, even for $N \sigma > 1$ \cite{vandyke:1974, song:steinrucken:2012}. This constitutes an interesting direction for future work. 

While the experimental applicability of these perturbation methods may be limited, they nevertheless provide a valuable \emph{qualitative} window into the effects of selection at many linked sites, and the exact nature of the selective corrections allows us to address a number of longstanding assumptions in the population genetics literature. Chief among these is the independent-sites assumption that is frequently used to model selection at individual sites along the genome. Somewhat surprisingly, we have demonstrated here that this assumption is valid not only in the purely neutral case, but also frequently through the first-order selective correction. At higher-orders, however, we start to obtain terms that depend on $NU$, and more generally, the variance in fitness maintained within the population. These terms represent corrections that arise solely from the interactions between the selected sites, and cannot be predicted from any single-locus theory.  

Of course, the standard assumption is not that linked sites evolve in this strictly independent fashion, but that they evolve independently at a reduced \emph{effective population size} $N_e$, which is supposed to encapsulate the effects of selection at neighboring sites \cite{hill:robertson:1966}. However, several recent studies have begun to challenge this assumption \cite{santiago:cabellero:1998, comeron:kreitman:2002, good:desai:2012}, often on the grounds that a different $N_e$ must be defined for every quantity we wish to predict. This shortcoming is apparent from our present analysis as well, and our analytical corrections provide an explicit demonstration. 

The effective population size is most commonly measured from the diversity at putatively neutral or synonymous sites. To lowest order in $Ns$, our analysis of the pairwise heterozygosity yields a corresponding effective population size
\begin{align}
\label{eq:diversity-ne}
N_e = N \left[ 1 - \frac{2 NU (Ns)^2}{9} \right] \, ,
\end{align}
which is reduced in the presence of selection as expected. To lowest order, this same $N_e$ correctly predicts the reduction of heterozygosity at selected sites as well. Alternatively, we could define $N_e$ by measuring the divergence (i.e., frequency of nucleotide substitutions) at various sites under selection, which depends on the fixation probability of a new mutant. To lowest order in $Ns$, this yields an effective population size 
\begin{align}
\label{eq:divergence-ne}
N_e = N \left[ 1 - \frac{2 NU (Ns)^2}{3} \right] \, , 
\end{align}
which is also reduced by selection at linked sites, but at slightly faster rate than in Eq. (\ref{eq:diversity-ne}). Thus, we require a different $N_e$ to account for linkage depending upon whether we wish to predict $\pi$, $p_\mathrm{fix}$, or some other sequence-based statistic. While the effective population size can still be used in the technical sense on a per-quantity basis, these results imply that its predictive or explanatory power is greatly reduced, and that the effects of linked selection are more complicated than a simple increase in genetic drift would suggest. 

In this way, the selective corrections obtained here can be extremely useful from a model-building standpoint, even when we wish to ultimately apply these models in regions where the perturbative approach breaks down. These corrections are straightforward to calculate for any quantity with a well-defined neutral limit, and because they are exact, any other model describing weakly selected mutations should recover these expressions as $Ns \to 0$. Many aspects of natural selection at the sequence level remain poorly understood, and exact results are few and far between. It is our hope that the methods outlined in the present work can be used as a stepping-stone to identify and evaluate those approximations which will lead to further progress on this important problem. 

\begin{acknowledgments}
This work was supported in part by the James S. McDonnell Foundation, the Alfred P. Sloan Foundation, and the Harvard Milton Fund. B.H.G. acknolwedges support from a National Science Foundation Graduate Research Fellowship. Simulations in this paper were performed on the Odyssey cluster supported by the Research Computing Group at Harvard University.
\end{acknowledgments}

\bibliographystyle{cbe}
\bibliography{evolution}

\newpage
\onecolumngrid
\begin{appendix}

\section{Finite site effects}

In the main text, we worked exclusively in the large-genome limit, where the number of sites was so large (and the per-site mutation rate so low) that we could focus on an intermediate asymptotic regime where back mutations could be neglected, and the mutation rate and distribution of fitness effects was independent of the previous mutations within a particular genome. In the present section we now consider what happens when we start to relax these assumptions. In particular, we assume that the genome has some finite size $L$ and that the per-site mutation rates are now sufficiently large that the scaled product $N\mu$ at each site is finite. For simplicity, we also assume a constant fitness effect for mutations at each site. This is similar to the model analyzed in \citet{woodcock:higgs:1996} and \citet{rouzine:etal:2003}. 

In this case, the population can still be partitioned according to the number of mutations $k=0,\ldots,L$ in each individual, but now we must account for the fact that the distribution of $k$-effects depends on $k$ in addition to the fitness. An individual with $k$ mutations can mutate at another site at rate $\mu(L-k)$, in which case $k \to k+1$. This individual can also experience a back-mutation at one of its $k$ mutated sites at rate $\mu k$, in which case $k \to k-1$. The total rate for one of these two events to happen is of course just $U = \mu L$.   Thus, the fitness classes $f(k)$ evolve according to the stochastic dynamics
\begin{align}
\frac{\partial f(k)}{\partial \tau} & = Ns(k-\overline{k}) f(k) + N \mu (L-k+1) f(k-1) + N \mu (k+1) f(k+1) - N \mu L f(k) \nonumber \\
	& \quad \quad  + \sum_{k'} \left[ \delta_{k k'} - f(k) \right] \sqrt{f(k')} \eta(k') \, . \label{eq:finite-sites-langevin}
\end{align}
In constrast to the $L \to \infty$ case analyzed in the main text, the behavior of the average profile $\langle f(k) \rangle$ in the neutral limit no longer degenerate, and we find that
\begin{align}
\label{eq:finite-average-profile}
\lim_{\tau \to \infty} \langle f(k,\tau) \rangle = {L \choose k} 2^{-L} \, ,
\end{align}
This is just a binomial distribution with mean $L/2$ and variance $L/4$, which is consistent with the intuition that the population at long times consists of individuals with independent and identically distributed mutations along the $L$ sites in the genome. Although there is no longer any infinite-width ``red-flag'' to suggest that fluctuations may play an important role here, the absence of any $N\mu$ dependence in Eq. (\ref{eq:finite-average-profile}) is suspicious, since we would expect that the typical width of the distribution should vanish as $N \mu \to 0$. 

Thus, we are lead to consider the behavior of the mean $\langle \overline{k} \rangle$ and the central moments $\M_m$ that we analyzed in the infinite-sites model. Using the dynamics in Eq. (\ref{eq:finite-sites-langevin}) it is straightforward to show that
\begin{align}
\frac{\partial \langle \overline{k} \rangle}{\partial \tau} & = Ns \M_2 + N\mu \left\langle \sum_k \left[ k(L-k+1) f(k-1) + k+1 f(k+1) - L f(k) \right] \right\rangle \\
	& = Ns \M_2 + N \mu L \left[ 1 - \frac{2 \overline{k}}{L} \right] \label{eq:finite-sites-mean}
\end{align}
so that in the neutral limit the population reaches mutation-reversion balance when the average number of mutations in each genome is $\langle \overline{k} \rangle = L/2$, just like the average profile in Eq. (\ref{eq:finite-average-profile}). We also see that this equilibrium point is reached on a timescale $t_{eq} \sim 1/2\mu$. For the central moments $\M_m$, it is straightforward to show that the equations become
\begin{align}
\frac{\partial \M_m}{\partial \tau} & = N\mu L \sum_{\ell=0}^{m-2} {m \choose \ell} \M_\ell + {m \choose 2} \M_{2,m-2} - m \M_m + Ns \left( \M_{m+1} - m \M_{2,m-1} \right) \nonumber \\
	& \quad \quad - 2 m N \mu \M_m - N \mu \sum_{\ell=0}^{m-2} {m \choose \ell} \left[ \M_{\ell+1} + \langle \overline{k} \M_\ell \rangle \right] \left[ 1 + (-1)^{m-\ell+1} \right]
\end{align}
where the last two terms arise from the $k$-dependent mutation rates. The first few orders of the moment hierarchy are given by
\begin{align}
\frac{\partial \M_2}{\partial \tau} & = N \mu L - \left[ 1 + 4 N \mu \right] \M_2 + Ns \M_3 \\
\frac{\partial \M_3}{\partial \tau} & = N \mu \left[ L - 2\langle \overline{k} \rangle \right] - 3 \left[ 1 + 2 N \mu \right] \M_3 + Ns \left[ \M_4 - 3 \M_{2,2} \right] \\
\frac{\partial \M_4}{\partial \tau} & = N\mu L + \left[ 6 N \mu L - 8 N \mu \right] \M_2 + 6 \M_{2,2} - 4 \left[ 1 + 2 N \mu \right] \M_4 + O(Ns) \\
\frac{\partial \M_{2,2}}{\partial \tau} & = 2 N \mu L \M_2 - \left[ 3+8N\mu \right] \M_{2,2} + \M_4 + O(Ns)
\end{align}
Thus, in the neutral limit, the actual variance in $k$ within the population is given by
\begin{align}
\M_2 = \frac{N \mu L}{1 + 4 N \mu}
\end{align}
which only approaches the deterministic value of $L/4$ when $N \mu \gg 1$. For smaller per-site mutation rates, the width of the fitness distribution can be much smaller than this, and for $N \mu \ll 1$ it approaches the infinite-sites limit $\M_2 = N \mu L$ that we analyzed in the main text. We can apply our perturbative approach to this moment hierarchy as well, which shows that when the mutants are weakly beneficial the equilibrium value of $\langle \overline{k} \rangle$ is given by
\begin{align}
\label{eq:perturbative-fixed-point}
\langle \overline{k} \rangle = \frac{L}{2} \left[ 1 + \frac{Ns}{1+4 N \mu} - \frac{2 Ns }{3} \left( \frac{Ns}{1+4 N \mu} \right)^2 \left( \frac{N \mu L + \frac{1}{2} + 4 N \mu + 16 (N \mu)^2}{3 + \frac{34}{3} N \mu + \frac{88}{3} (N \mu)^2 + \frac{64}{3} (N \mu)^3 } \right) \right] + O(Ns)^4
\end{align}
Following \citet{woodcock:higgs:1996} and \citet{goyal:etal:2012}, we define $\epsilonc$ to be the fraction of all the possible mutations that are beneficial at this equilibrium point, or
\begin{align}
\epsilonc = \frac{(L-\langle \overline{k} \rangle)\mu}{L \mu} = 1 - \frac{\langle \overline{k}\rangle}{L} 
\end{align}
This allows us to rewrite Eq. (\ref{eq:perturbative-fixed-point}) in terms of this critical ratio as
\begin{align}
\epsilonc = \frac{1}{2} \left[ 1 - \frac{Ns}{1+4 N \mu} + \frac{2 Ns }{3} \left( \frac{Ns}{1+4 N \mu} \right)^2 \left( \frac{N \mu L + \frac{1}{2} + 4 N \mu + 16 (N \mu)^2}{3 + \frac{34}{3} N \mu + \frac{88}{3} (N \mu)^2 + \frac{64}{3} (N \mu)^3 } \right) \right] + O(Ns)^4
\end{align}
In the limit that $N \mu \to 0$ and $L \to \infty$ with $NU = N \mu L$ fixed, we recover the infinite sites preduction
\begin{align}
\epsilonc = \frac{1}{2} \left[ 1 - Ns + \frac{(Ns)^3}{3} + \frac{2 NU (Ns)^3}{3} \right] + O(Ns)^4
\end{align}
which provides a more accurate expression for the critical fraction as $Ns \to 0$ compared to the corresponding expression in \citet{goyal:etal:2012}.

\section{Stochastic Calculus}

In this section, we outline the stochastic (It\^{o}) calculus that is used to derive moment equations from the stochastic dynamics in Eqs. (\ref{eq:diffusion}), (\ref{eq:langevin}), and (\ref{eq:neutral-langevin}). For concretness, we will restrict our attention to the neutral dynamics in Eq. (\ref{eq:neutral-langevin}), but these results will apply more generally. Let $\phi(\{f_k\})$ be some arbitrary function of the fitness classes, $f_k(t)$. We wish to find an expression for the time-evolution of the mean $\langle \phi(\{f_k\}) \rangle$ using the definition
\begin{align}
\frac{\partial \langle \phi(\{ f_k \})}{\partial t} & = \lim_{\delta t \to 0} \frac{\langle \phi(\{ f_k(t+\delta t) \} \rangle - \langle \phi(\{ f_k(t) \}) \rangle}{\delta t} \, .
\end{align}
The dynamics in Eq. (\ref{eq:neutral-langevin}) are essentially just a shorthand notation for calculating $f_k(t+\delta t)$ conditioned on $f_k(t)$. Taking care to note that the drift term in Eq. (\ref{eq:neutral-langevin}) is of the It\^{o} form, the dynamics in Eq. (\ref{eq:neutral-langevin}) imply that 
\begin{align}
f_k(t+\delta t) & = f_k(t) + \delta t \left[ U f(k-1,t) + U f(k,t) \right] \nonumber \\
	& \quad \quad + \sqrt{\delta t} \left[ \frac{1}{\sqrt{N}} \sum_{k'} \left[ \delta_{k k'} - f(k,t) \right] \sqrt{f(k',t)} \eta(k',t) \right] \, .
\end{align} 
The value of $\phi(\{f_k(t+\delta t)\})$ can then be found by Taylor expansion in powers of $\delta  t$. If $\phi(\{f_k\})$ is just a singleton function $\phi(\{f_k\}) = f_k(t)$, then 
\begin{align}
\langle \phi(\{f_k(t+\delta t)\}) \rangle&  = \langle f_k(t) \rangle + \delta t \left\langle U f(k-1,t) + U f(k,t) \right\rangle \\
	& = \langle f_k(t) \rangle + \delta t \left\langle \left( \frac{\partial f_k}{\partial t} \right)_\mathrm{det} \right\rangle \, ,
\end{align}
where $(\partial f_k/\partial t )_\mathrm{det}$ is simply the deterministic part of the dynamics in Eq. (\ref{eq:neutral-langevin}). On the other hand, if $\phi(\{f_k\})$ is a pairwise product of the form
\begin{align}
\phi(\{f_k\}) = f_{k_1}(t) f_{k_2}(t) \, ,
\end{align}
then things start to become more complicated. Organizing all the terms in powers of $\delta t$, we see that
\begin{align}
\label{eq:product-dynamics}
f_{k_1}(t+\delta t) f_{k_2}(t+\delta t) & = f_{k_1} f_{k_2} + \delta t \left[ \left( U f_{k_1-1} - U f_{k_1} \right) f_{k_2} + f_{k_1} \left( U f_{k_2-1} - U f_{k_2} \right) \right. \nonumber \\
	& \quad \quad \left. + \frac{1}{N} \sum_{j_1,j_2} \left[ \delta_{k_1,j_1} - f_{k_1} \right] \left[ \delta_{k_2,j_2} - f_{k_2} \right] \sqrt{f_{j_1} f_{j_2}} \eta_{j_1} \eta_{j_2} \right] + \sqrt{\delta t} \, O(\eta) \, .
\end{align}
The term proportional to $\sqrt{\delta t}$ is linear in $\eta$, so upon averaging this term vanishes. The term proportional to $\delta t$ also contains a term involving $\eta$, but this time as a quadratic function rather than a linear function, which yields
\begin{align}
\left\langle \sum_{j_1,j_2} \left[ \delta_{k_1,j_1} - f_{k_1} \right] \left[ \delta_{k_2,j_2} - f_{k_2} \right] \sqrt{f_{j_1} f_{j_2}} \eta_{j_1} \eta_{j_2}  \right\rangle & = \left\langle \sum_{j} \left[ \delta_{k_1,j} - f_{k_1} \right] \left[ \delta_{k_2,j} - f_{k_2} \right] f_j  \right\rangle \\
	& = \left[ \delta_{k_1,k_2} - f_{k_2} \right] f_{k_1} \, ,
\end{align}
where we have used the fact that the $\eta_k$ are uncorrelated for different $k$. Thus, taking the average of Eq. (\ref{eq:product-dynamics}), we obtain
\begin{align}
\frac{\partial \langle f(k_1,t) f(k_2,t) \rangle}{\partial t} & = \left\langle f(k_1,t) \left( \frac{\partial f(k_2,t)}{\partial t} \right)_\mathrm{det} \right\rangle + \left\langle \left( \frac{\partial f(k_1,t)}{\partial t}\right)_{\mathrm{det}} f(k_2,t) \right\rangle \nonumber \\
	& \quad \quad  + \frac{1}{N} \langle f(k_1,t) \star f(k_2,t) \rangle \, , \label{eq:pairwise-product-rule}
\end{align}
where we have defined a new operation $\star$ such that
\begin{align}
f(k_1,t) \star f(k_2,t) \equiv \left[ \delta_{k_1,k_2} - f_{k_2} \right] f_{k_1} \, .
\end{align}
This is the generalized product rule of the It\^{o} calculus, which arises from the combination of two $\sqrt{\delta t}$ terms in the expansion of the Langevin equation. More complicated functions of the $f_k$ can be analyzed recursively with the help of the sum and product rules
\begin{align}
\frac{\partial \langle \phi(f) + \psi(f)\rangle}{\partial t} = & = \left\langle\frac{\partial \phi(f)}{\partial t} \right\rangle + \left\langle \frac{\partial \psi(f)}{\partial t}  \right\rangle 
\end{align}
and
\begin{align}
\label{eq:product-rule}
\frac{\partial \langle \phi(f) \psi(f)\rangle}{\partial t} = & = \left\langle \frac{\partial \phi(f)}{\partial t} \psi(f) \right\rangle + \left\langle \phi(f) \frac{\partial \psi(f)}{\partial t} \right\rangle + \frac{1}{N} \langle \phi(f) \star \psi(f) \rangle \, ,
\end{align}
where $\star$ is defined in terms of the underlying fitness classes $f_k$ and satisfies additivity and the distributive property. As an example, we have
\begin{align}
\overline{k} \star f(k,t) & = \sum_{k'} k' \left[ f(k,t) \star f(k',t) \right] \\
	& = \sum_{k'} k' \left[ \delta(k-k') - f(k) \right] f(k') \\
	& = (k-\overline{k}) f(k,t) \, ,
\end{align}
and 
\begin{align}
\overline{k} \star \overline{k} & = \sum_{k} k [\overline{k} \star f(k,t)] \\
	& = \sum_{k} k(k-\overline{k}) f(k,t) \\
	& = \sum_{k} (k-\overline{k})^2 f(k,t) \, .
\end{align}
These rules can be used to rapidly generate equations of motion for the generalized moments $\M_{\vec{m}}$ and $\F_{\vec{m};\vec{n}}$ analyzed in the text (see Appendices C and D). 

\section{Central Moments}

In this section, we use the rules of the stochastic calculus in Appendix B to derive equations of motion for the mean ``fitness'' 
\begin{align}
\overline{k} = \int dk \, k f(k,t) 
\end{align} 
and the central moments 
\begin{align}
\M_m = \left\langle \int dk \, (k-\overline{k})^m f(k,t) \right\rangle
\end{align}
Without loss of generality, we will restrict our attention to the full dynamics in Eq. (\ref{eq:k-langevin}), where $k$ is a continuous variable proportional to the absolute fitness. Directly from the Langevin equation, we can see that
\begin{align}
\frac{\partial \langle \overline{k} \rangle}{\partial \tau} & = \int dk \, k \left\langle \frac{f(k,\tau)}{\partial \tau} \right\rangle \\
	& = \int dk \, k \left\langle  NU \int d(\deltak) \, \rho(\deltak) \left[ f(k-\deltak) - f(k) \right] + Ns (k-\overline{k}) f(k) \right\rangle \\
	& = NU \langle \deltak \rangle + Ns \cdot \M_2 \, .
\end{align}
For the central moments $\M_m$, we can use the product rule in Eq. (\ref{eq:product-rule}) to show that
\begin{align}
\frac{\partial \M_m}{\partial \tau} & = \left\langle \int dk \, (k-\overline{k})^m \frac{\partial f(k)}{\partial t} - m \left[ \int dk \,  (k-\overline{k})^{m-1} f(k) \right] \frac{\partial \overline{k}}{\partial \tau} \right\rangle \nonumber \\
	& \quad \quad + \left\langle - m \int dk \, (k-\overline{k})^{m-1} \left[ \overline{k} \star f(k) \right] + { m \choose 2} \left[ \int dk \, (k-\overline{k})^{m-2} f(k) \right]  \overline{k} \star \overline{k} \right\rangle \, ,
\end{align}
or
\begin{align}
\frac{\partial \M_m}{\partial \tau} & = NU \sum_{\ell=0}^{m-2} {m \choose \ell} \langle \delta k^{m-\ell} \rangle \M_\ell + {m \choose 2} \M_{2,m-2} - m \M_m + Ns \left( \M_{m+1} - m \M_{2,m-1} \right) \, .
\end{align}
In terms of the rescaled moments $\Mtilde_m = \M_m / (NU)^{m/2}$, this can be rewritten in the form
\begin{align}
\frac{\partial \Mtilde_m}{\partial \tau} & = \sum_{\ell=0}^{m-2} {m \choose 2 + \ell} \left[ \frac{\left\langle \deltak^{2+\ell} \right\rangle}{(NU)^{\ell/2}} \right] \Mtilde_{m-\ell-2} + {m \choose 2} \Mtilde_{2,{m-2}} - m \Mtilde_m \nonumber \\
	& \quad \quad + N\sigma \left[ \Mtilde_{m+1} - m \Mtilde_{2,m-1} \right] \, ,
\end{align}
where $N\sigma \equiv NU(Ns)^2$. Thus, in the limit that $NU \to \infty$, this reduces to
\begin{align}
\frac{\partial \Mtilde_m}{\partial \tau} & = {m \choose 2} \langle \deltak^2 \rangle \Mtilde_{m-2} + {m \choose 2} \Mtilde_{2,{m-2}} - m \Mtilde_m \nonumber + N\sigma \left[ \Mtilde_{m+1} - m \Mtilde_{2,m-1} \right] \, ,
\end{align}
which is independent of $NU$ and depends on the distribution of fitness effects only through the second moment $\langle \deltak^2 \rangle$. 

In order to obtain equations of motion for the generalized products $M_{m_1,\ldots,m_J}$, we can again appeal to the product rule in Eq. (\ref{eq:product-rule}) which shows that (with some abuse of notation)
\begin{align}
\frac{\partial \M_{m,n}}{\partial \tau} & = \left\langle \frac{\partial \M_m}{\partial \tau} \M_n  \right\rangle + \left\langle \M_m \frac{\partial \M_n}{\partial \tau} \right\rangle + \langle \M_m \star \M_n \rangle \, .
\end{align}
This requires us to compute
\begin{align}
\overline{k} \star \M_m & = \int dk \, (k - \overline{k})^m [\overline{k} \star f(k,t) ] - m \M_{m-1} ( \overline{k} \star \overline{k} ) \\
	& = \M_{m+1} - m \M_{2,m-1} 
\end{align}
and
\begin{align}
\M_{m} \star \M_{n} & = \int dk_1 \, dk_2 \, (k_1-\overline{k})^m (k_2-\overline{k}) [f(k_1) \star f(k_2)] - m \M_{m-1} \int dk \, (k-\overline{k})^n [ \overline{k} \star f(k) ] \nonumber \\
	& \quad \quad - n \M_{n-1}  \int dk \, (k-\overline{k})^m [\overline{k} \star f(k)] + m n (\overline{k} \star \overline{k} ) \M_{m-1} \M_{n-1} \\
	& = \M_{m+n} - \M_{m,n}- m \M_{m-1,n+1} - n \M_{m+1,n-1} + m n \M_{2,m-1,n-1} \, . 
\end{align}
The equations of motion for generalized products with more than two terms follow from the product rule in Eq. (\ref{eq:product-rule}). 

\section{Focal Lineage Moments}

In this section, we use the rules of the stochastic calculus in Appendix B to derive equations of motion for the focal lineage moments
\begin{align}
\F_m = \left\langle \int dk \, (k-\overline{k})^m f_1(k,t) \right\rangle
\end{align}
discussed in the main text. Starting from the product rule in Eq. (\ref{eq:product-rule}), we have
\begin{align}
\frac{\partial \F_m}{\partial \tau} & = \left\langle \int dk \, (k-\overline{k})^m \frac{\partial f_1(k)}{\partial t} - m \left[ \int dk \, (k-\overline{k})^{m-1} f_1(k) \right] \frac{\partial \overline{k}}{\partial \tau} \right\rangle \nonumber \\
	& \quad \quad + \left\langle - m \int dk \, (k-\overline{k})^{m-1} \left[ \overline{k} \star f_1(k) \right] + { m \choose 2} \left[ \int dk \, (k-\overline{k})^{m-2} f_1(k) \right]  \overline{k} \star \overline{k} \right\rangle \, .
\end{align}
In order to progress further, we must extend the $\star$ operation to the case where the fitness classes $f_i(k,t)$ are labeled according to which lineage they descend from. It is a straightforward matter to show that
\begin{align}
f_{i_1}(k_1,t) \star f_{i_2}(k_2,t) = \left[ \delta_{i_1 i_2} \delta(k_1-k_2) - f_{i_2}(k_2) \right] f_{i_1}(k_1) \, .
\end{align}
Thus, we have
\begin{align}
\overline{k} \star f_1(k,t) & = \sum_i \int dk' \,  k' [f_i(k',t) \star f_1(k,t)] \\
	& = \sum_i \int dk' \, k' \left[ \delta(k-k') \delta_{i1} - f_1(k,t) \right] f_i(k',t) \\
   & =  (k-\overline{k}) f_1(k,t) \, ,
\end{align}
and we can immediately conclude that
\begin{align}
\frac{\partial \F_m}{\partial \tau} & = NU \sum_{\ell=0}^{m-2} {m \choose \ell} \langle \deltak^{m-\ell} \rangle \F_\ell + {m \choose 2} \F_{m-2;2}  - m \F_m + Ns \left( \F_{m+1} - m \F_{m-1;2} \right) \, .
\end{align}
In order to derive the equations of motion for the general products $\F_{\vec{m};\vec{n}}$ considered in the text, we simply need to calculate $\star$ products of the form $\F_m \star \M_n$ and $\F_m \star \F_n$. Starting from
\begin{align}
\overline{k} \star \F_m & = - m \F_{m-1} (\overline{k} \star \overline{k}) + \int dk \, (k-\overline{k})^m [\overline{k} \star f_1(k,t) ] \\
	& = \F_{m+1} - m \F_{m-1;2} \, ,
\end{align}
we can easily see show that
\begin{align}
\F_m \star \M_n & = \sum_i \int dk_1 \, dk_2 \, (k_1 - \overline{k})^m (k_2 - \overline{k})^n [f_1(k_1) \star f_i(k_2)] \nonumber \\
	& \quad \quad - m \F_{m-1} \sum_i \int dk \, (k-\overline{k})^m [\overline{k} \star f_i(k)] - n \M_{n-1} \int dk \, (k - \overline{k})^m [ \overline{k} \star f_1(k) ] \nonumber \\
	& \quad \quad + m n \F_{m-1} \M_{m-1} [\overline{k} \star \overline{k} ] \\
	& = \F_{m+n} - \F_{m;n} - m \F_{m-1;n+1} - n \F_{m+1;n-1} + m n \F_{m-1;n-1,2} 
\end{align}
and
\begin{align}
\F_m \star \F_n & = \int dk_1 \, dk_2 \, (k_1-\overline{k})^m (k_2 - \overline{k})^n [f_1(k_1) \star f_1(k_2) ] - m \F_{m-1} \int dk \, (k-\overline{k})^n [\overline{k} \star f_1(k)] \nonumber \\
	& \quad \quad - n \F_{n-1} \int dk \, (k-\overline{k})^m [\overline{k} \star f_1(k)] + m n \F_{m-1} \F_{n-1} [\overline{k} \star \overline{k}] \\
	& = \F_{m+n} - \F_{m,n} - m \F_{m-1,n+1} - n \F_{m+1,n-1} + m n \F_{m-1,n-1;2} \, . 
\end{align}
The equations of motion for generalized products with more than two terms follow from the product rule in Eq. (\ref{eq:product-rule}). 


\end{appendix}
\end{document}